\documentstyle[preprint,pre,aps,eqsecnum]{revtex}
\begin{document}

\title{Comparison of Canonical and Grand Canonical Models for
selected multifragmentation data}
 
\author{C. B. Das$^1$, S. Das Gupta$^1$, X. D. Liu$^2$ and M. B. Tsang$^2$}

\address{
$^1$Physics Department, McGill University, Montr{\'e}al, Canada H3A 2T8\\
$^2$National Superconducting Cyclotron Laboratory, East Lansing, MI 48824,
USA}

\date{\today}

\maketitle

\begin{abstract}
Calculations for a set
of nuclear multifragmentation data are made using a Canonical
and a Grand Canonical Model.  The physics assumptions are identical
but the Canonical Model has an exact number of particles, whereas,
the Grand Canonical Model has a varying number of particles, hence,
is less exact.  Interesting differences are found.
\end{abstract}


\section{Introduction}
In experiments whose goals were to investigate the role of isospin
in fragment yields \cite{Xu00}, the following interesting features have been
observed.  If we compare the central collisions of two heavy ion systems, 1
and 2, which are similar in all aspects of the reactions except for the
neutron and proton composition, the isotope yield ratios, 
Y$_2$(n,z)/Y$_1$(n, z),
where Y$_i$(n,z) is the yield of the isotope with neutron number n, and proton
number z, from reaction $i$, is found to exhibit an exponential relationship
as a function of n and z \cite{Xu00,Tsang}
\begin{eqnarray}
Y_2(n,z)/Y_1(n,z)=Cexp(\alpha_nn+\alpha_pz)
\end{eqnarray}	 		
where $C$ is an overall normalization constant and $\alpha_n$ and $\alpha_p$ 
are fitting
parameters. This phenomenon is termed isoscaling, a strong evidence that
the processes are statistical.

Related to isoscaling is the exponential dependence of the mirror nuclei
ratios on the binding energy. In Figure 1, 
the ratios of yields of mirror nuclei: $Y_i(t)/Y_i(^3He)$, $Y_i(^7Li)/
Y_i(^7Be)$ and $Y_i(^{11}B)/Y_i(^{11}C)$ for central collisions of
$^{124}Sn+^{124}Sn$ (solid points) and $^{112}Sn+^{112}Sn$ (open points)
at 50 MeV per nucleon are plotted
as a function of the binding energy difference, $\Delta E_B$.
These ratios fall approximately on an exponential.
Many statistical models such as the
grand canonical model \cite{Randrup,Albergo} of
multifragmentation predict
both the isoscaling and mirror-nuclei ratio dependence.

Experimental evidence suggests that multifragmentation occurs when the
heated matter expands to density about 1/3 of nuclear matter density
\cite{Bowman} and the time scale for the
emission of fragments is short, between 50 to 100 fm/c \cite{Cornell}.
Most successful statistical models that describe multifragmentation data
assume a freeze-out volume at which
composite yields are to be calculated entirely according to phase-space
\cite{Bondorf,Gross}.
If the dissociating system is very large, then grand canonical
simplification
can be employed \cite{Randrup,Albergo}.  According to this model, the average
number of composites
with neutron number $i$ and proton number $j$ is
\begin{eqnarray}
<n_{i,j}>=\exp(\beta(i\mu_n+j\mu_p))\omega_{i,j}
\end{eqnarray}
where $\mu_n,\mu_p$ are neutron and proton chemical potentials and
\begin{equation}
\omega_{i,j}=\frac{V}{h^3}(2\pi (i+j)mT)^{3/2}(2s+1)z_{int}
\exp(\beta E_B) \nonumber
\end{equation}
is the partition function of one composite. $\beta$ is the inverse
temperature.
For mirror nuclei: $i=k+1,k$ and $j=k,k+1$ we should simply have
\begin{eqnarray}
\frac{<n_{k+1,k}>}{<n_{k,k+1}>}=\exp(\beta\mu_n-\beta\mu_p)\exp(\beta\Delta
 E_B)
\end{eqnarray}
Thus the log of the ratios of the yield will be linear with respect
to $\Delta E_B$ which is approximately obeyed by data.
However a more close inspection raises another issue.

According to Eq. 1.4, one can deduce the value of $\beta=1/T$ from the slope of the line.
Indeed for the lines drawn in Fig.1, the temperature $T$ is less than
2 MeV.  For such a low temperature, the model of
simultaneous breakup model [7,8] 
should not be appropriate. In addition, such low values
are in direct contradictions with temperature measurements
obtained from isotope yield ratios.
The isotope yield temperature is about 5 MeV for the Sn+Sn systems
\cite{Xu01,Kunde}.
To resolve the discrepancies between temperatures observed,
it is necessary to explore details of the exponential behaviour of the
mirror nuclei.

\section{Canonical {\it vs.} Grand Canonical Models}
In recent years, the grand canonical model has been replaced by a
canonical model.  The physics assumptions are still the same but
we no longer have to assume that the system is large.  This is a
technical advancement; the details have already been described in several
places \cite {Dasgupta,Majumder,Bhattacharyya} so we will not 
repeat these here.  The model has been used
to fit the isotope data \cite{Majumder,tsang}. Surprisingly,
isoscaling which follows naturally from the Grand Canonical model, emerges
also in canonical model \cite{tsang}.
In this article, we will investigate why certain results from the canonical
model
resemble those from the grand canonical model and what are the differences.
We will also investigate the
relation between the canonical temperature and the temperature obtained
based on the simpler grand canonical rules.

The yield of the composite which has $k+1$ neutrons and $k$ protons is given
in the canonical model by
\begin{eqnarray}
<n_{k+1,k}>=\omega_{k+1,k}\frac{Q_{N-k-1,Z-k}}{Q_{N,Z}}
\end{eqnarray}
Here $N,Z$ refer to the number of neutrons and protons of the 
disintegrating system. $Q_{N,Z}$ is the canonical partition function of
this system.  Similarly, $Q_{N-k-1,Z-k}$ is the canonical partition function
of the residue system which has $N-k-1$ neutrons and $Z-k$ protons.
The ratio of the yields in the canonical model is then given by
\begin{eqnarray}
\frac{<n_{k+1,k}>}{<n_{k,k+1}>}=\frac{\omega_{k+1,k}}{\omega_{k,k+1}}\times
\frac{Q_{N-k-1,Z-k}}{Q_{N-k,Z-k-1}}
\end{eqnarray}
The first factor leads to  $\exp(\beta\Delta E_B)$.  We note
in passing that for mirror nuclei $\Delta E_B=\Delta E_C$,
the change in Coulomb energy.  If we assume a uniformly charged sphere,
then $\Delta E_c=\frac{3}{5}\frac {e^2}{R_0a^{1/3}}[(z+1)^2-z^2]=.72a^{2/3}$
MeV where $a$ is the composite mass number.  
For light nuclei $0.72a^{2/3}$ MeV does not fit the data very well.
We note for later use that $ 0.235a$ MeV fits $\Delta E_B$ between
$a=7$ and $a=15$ better.

The exact expression for the canonical partition function $Q_{N,Z}$ 
used in \cite {tsang} does not allow us to investigate easily the 
features we want to study. Since the ratios
are very simple in the grand canonical ensemble and since there is a
connection between grand canonical partition function $Z_{gr}(\lambda_n,
\lambda_p)$ and the canonical partition function $Q_{N,Z}$, we find it
convenient to exploit this relation.  In the present problem, the
grand canonical partition function is given by
\begin{eqnarray}
Z_{gr}(\lambda_n,\lambda_p)=\sum_{k,l,n_{k,l}}
 e^{(k\lambda_n+l\lambda_p)n_{kl}}
\times\frac{\omega_{k,l}
^{n_{k,l}}}{n_{k,l}!}
\end{eqnarray}
The expression for $logZ_{gr}(\lambda_n,\lambda_p)$ is
\begin{eqnarray}
logZ_{gr}(\lambda_n,\lambda_p)=\sum_{k,l}\exp(k\lambda_n+l\lambda_p)\times
\omega_{k,l}
\end{eqnarray}
The canonical partition function can be obtained from $Z_{gr}$ by Laplace
inverse:
\begin{eqnarray}
Q_{N,Z}=\frac{1}{(2\pi)^2}\int_{-\pi}^\pi\int_{-\pi}^\pi
e^{-(\lambda_n+i\tilde{\lambda}_n)N}e^{-(\lambda_p+i\tilde{\lambda}_p)Z}
e^{logZ_{gr}(\lambda_n+i\tilde{\lambda}_n,\lambda_p+i\tilde{\lambda}_p)}
d\tilde{\lambda}_nd\tilde{\lambda}_p
\end{eqnarray}
While this expression is true for any $\lambda_n$ and $\lambda_p$, the
saddle-point approximation consists in choosing the values of $\lambda_n$ and
$\lambda_p$ such that the kernel maximizes at $\tilde{\lambda}_n=0$ and
$\tilde{\lambda}_p=0$ and making a Gaussian approximation for the integrand
around this maximum.  The result is
\begin{eqnarray}
Q_{N,Z}\approx
e^{-(\lambda_nN+\lambda_pZ)}e^{logZ_{gr}(\lambda_n,\lambda_p)}/
(2\pi *|det|^{1/2})
\end{eqnarray}
where the values of $\lambda_n$ and $\lambda_z$ are such that the average
numbers of neutrons and protons as obtained from the grand canonical
ensemble
are $N$ and $Z$, i.e.,
\begin{eqnarray}
N=\frac{\partial logZ_{gr}(\lambda_n,\lambda_p)}{\partial \lambda_n};
\end{eqnarray}
\begin{eqnarray}
Z=\frac{\partial logZ_{gr}(\lambda_n,\lambda_p)}{\partial \lambda_p}
\end{eqnarray}
The elements of the determinant are given by:
$a_{1,1}=\frac{\partial^2log Z_{gr}}{\partial^2\lambda_n}, a_{1,2}=a_{2,1}=
\frac{\partial^2log Z_{gr}}{\partial\lambda_n\partial\lambda_p}$ and
$a_{2,2}=\frac{\partial^2log Z_{gr}}{\partial^2\lambda_p}$.

Eq. (2.2) now takes the form
\begin{eqnarray}
\frac{<n_{k+1,k}>}{<n_{k,k+1}>}\approx e^{\beta\Delta E_B}\times\frac{
e^{-(\lambda_n(N-k-1)+\lambda_p(Z-k))+logZ_{gr}(\lambda_n,\lambda_p)}}{
e^{-(\lambda_n'(N-k)+\lambda_p'(Z-k-1))+logZ_{gr}(\lambda_n',\lambda_p')}}
\end{eqnarray}
Here we have omitted the ratios of the determinants $|det|^{1/2}$ because 
their effects will be negligible.
Eq. (2.9) will reduce to the standard grand canonical result if
we set $\lambda_n=\lambda_n'; \lambda_p=\lambda_p'$ and take
these values from a system which has the average number of neutrons
to be $N$ ( rather than $N-k-1$ to get $\lambda_n$ and $N-k$ to get
$\lambda_n'$ ) and the average number of protons to be $Z$ ( rather than
$Z-k$ to obtain $\lambda_p$ and $Z-k-1$ to obtain $\lambda_p'$).

For a better estimate, let us write $\lambda_n'=\lambda_n+\Delta\lambda_n;
 \lambda_p'=\lambda_p+\Delta\lambda_p$.  Assuming lowest order expansion
is valid we can get (depending upon whether we expand $logZ_{gr}(\lambda_n,
\lambda_p)$ in terms of $logZ_{gr}(\lambda_n',\lambda_p')$ or vice versa):

$e^{(logZ_{gr}(\lambda_n,
\lambda_p)-logZ_{gr}(\lambda_n',\lambda_p'))}=e^{-\Delta\lambda_n(N-k-1)-
\Delta\lambda_p(Z-k)}$ or
$e^{-\Delta\lambda_n(N-k)-\Delta\lambda_p(Z-k-1)}$.

Eq. (2.9) can be reduced to
$\frac{<n_{k+1,k}>}{<n_{k,k+1}>}\approx e^{\beta\Delta E_B} \times
e^{\lambda_n-\lambda_p}
\approx e^{\beta\Delta E_B}\times e^{\lambda_n'-\lambda_p'}$

We will use
\begin{eqnarray}
\frac{<n_{k+1,k}>}{<n_{k,k+1}>}\approx e^{\beta\Delta E_B}\times e^{(\lambda_n+
\lambda_n'-\lambda_p-\lambda_p')/2}
\end{eqnarray}
Eq. (2.10) looks just like a grand canonical result but with an important
difference.  In the usual grand canonical model $\lambda_n,\lambda_p$
would be calculated just once, from eqs. (2.7) and (2.8) where 
$N$ and $Z$ are the neutron and proton numbers
of the disintegrating system.  By contrast, $\lambda_n,\lambda_p$ etc. of
eq. (2.10) are calculated from eqs (2.7) and (2.8) for each $k$
and the left hand sides
of  eqs.(2.7) and (2.8) are given by $N-k-1$ and $Z-k$ respectively.
The quantity $\lambda_n-\lambda_p$ etc. increases with $k$ with the result
that if we try to interpret the canonical results within a usual grand
canonical framework one ends up with a larger $\beta$, that is, a lower $T$.
This is demonstrated in Fig. 2 where it is shown that although the
temperature
used for the canonical calculation (hence the true temperature) is 5 MeV,
deducing the temperature from the slope of the mirror isotope yield
ratios (as one would do in a grand canonical
formalism) one would arrive at a significantly lower temperature.
The best fit (solid line) to the calculated values from the canonical 
model (solid points) yield a temperature of 3.395 MeV.
In the figure we also show that the approximation of Eq.(2.10) 
as shown by the star points, works quite well.

The dependence of $\lambda_n-\lambda_p$ on $k$ and $N,Z$ where $2k+1$
is the mass number of the emitted particle and $N,Z$ gives the size
of the emitting system can be pinned down further.  Let
$\Lambda_N,\Lambda_P$
be the fugacities of the system $N,Z$.  We will write $\lambda_n=
\Lambda_N+d\lambda_N$ and $\lambda_p=\Lambda_P+d\lambda_P$.  We then
have
$N-k-1=\sum i\omega_{i,j}\exp(i\lambda_n+j\lambda_p)$.  Expressing
$\lambda_n,\lambda_p$ in terms of $\Lambda_N,\Lambda_P$
and approximating $\exp(d\Lambda_P)\approx
(1+d\Lambda_P)$ etc. we get $-k-1=Ad\Lambda_N+Bd\Lambda_P$
where $A$ and $B$ are constants:
$A=\sum i^2\omega_{i,j}\exp(i\Lambda_N+j\Lambda_P)$ and $B=
\sum ij\omega_{i,j}\exp(i\Lambda_N+j\Lambda_P)$.  Similarly starting
from $N-k=\sum j\omega_{i,j}\exp(i\lambda_n+j\lambda_p)$ and expanding
as above we get $-k=Bd\Lambda_N+Cd\Lambda_P$ where $C=
\sum j^2\omega_{i,j}\exp(i\Lambda_N+j\Lambda_P)$.  One can now
express $d\Lambda_N, d\Lambda_P$ in terms of the constants $A, B$
and $C$.  We get
\begin{eqnarray}
\lambda_n-\lambda_p=\Lambda_N-\Lambda_P+\frac{C-A}{B^2-AC}k+
\frac{C+B}{B^2-AC}
\end{eqnarray}
We can do a similar analysis for $\lambda_n'-\lambda_p'$.  Finally
we get (compare eq.(2.10))
\begin{eqnarray}
(\lambda_n+\lambda_n'-\lambda_p-\lambda_p')/2=\Lambda_N-\Lambda_P+
\frac{1}{2}\frac{C-A}{B^2-AC}(2k+1)
\end{eqnarray}
Eq.(2.12) says that the correction grows like $2k+1=a$, the mass
of the composite.  The correction
would diminish as the disintegrating system ($N,Z$) grows.  The
constants $A, B$ and $C$ are positive definite and each will become
larger and larger as the disintegrating system becomes larger.  The
correction would disappear in the thermodynamic limit.  The actual
values of the constants $A, B$ and $C$ for a finite system depend
on many factors: the symmetry energy, the Coulomb energies and $N,Z$
of the disintegrating system.

\section{The Albergo Temperature}
The Albergo formula \cite {Albergo} has often been used to extract
a temperature from experimental data.  The formula is exact if
the following two assumptions are valid: (1) the populations
of various states are given by the grand canonical model and (2)
the secondary decays which will alter these primary populations can
be neglected.  Define a ratio $R$
\begin{eqnarray}
R=\frac{Y(A_i,Z_i)/Y(A_i+1,Z_i)}{Y(A_j,Z_j)/Y(A_j+1,Z_j)}
\end{eqnarray}
where the $Y$'s are the yields in the ground state.
Then, the temperature is given by
\begin{eqnarray}
T=\frac{B}{ln(sR)}
\end{eqnarray}
where $B$ is related to binding energies and $s$ to the ground state spins:
\begin{eqnarray}
B=BE(A_i,Z_i)-BE(A_i+1,Z_i)-BE(A_j,Z_j)+BE(A_j+1,Z_j)
\end{eqnarray}
\begin{eqnarray}
s=\frac{[2S(A_j,Z_j)+1]/[2S(A_j+1,Z_j)+1]}
{[2S(A_i,Z_i)+1]/[2S(A_i+1,Z_i)+1]}
\end{eqnarray}
Even if the grand canonical model is exact, the change of populations
due to secondary decays can cause eq.(3.2) to give significantly
different temperatures from the true grand canonical temperature.  
This was studied in
detail in \cite{Tsang97}.  It was shown that for large values of $B$
(eq.(3.3)), the difference between apparent temperature and the true
grand canonical temperature decreases.  This suggests that while
using the Albergo formula to deduce a temperature from experimental
data, it is advisable to use pairs that will lead to a large value of $B$.

Our objective here is different and is complimentary to the study made
in \cite{Tsang97}.  The canonical model is obviously more rigorous than
the grand canonical model. However, if canonical values for $R$ are used,
eq. (3.2) is no longer strictly correct.  Using the primary yields, we explore the differences between the 
deduced temperatures from
eq.(3.2) compared to the actual temperature used in a canonical model.
This is shown in Fig.3, we find that the
errors decrease with increasing $B$.  
The inset in Figure 3 shows the deviation of the canonical
Albergo temperature for B greater than 10 MeV. Most of the predicted
temperatures are slightly lower than the actual temperature of 5 MeV. 
The deviations arise from the differences between isotope
yields predicted by the canonical and grand canonical models
Not surprisingly, the conclusions of \cite {Tsang97} can be applied here.

\section{The scaling Law}

The last quantity we want to investigate is a ratio of two ratios:
$\frac{<n_{l+m,k}>_2}{<n_{l+m,k}>_1}\div \frac{<n_{l,k}>_2}{<n_{l,k}>_1}$
and see if this falls on an exponential as in the grand canonical ensemble.
Here the subscripts 1 and 2 refer to two systems : ( for example: 2 refers
to central collisions of $^{124}$Sn+$^{124}$Sn and 1 to central collisions
of $^{112}$Sn+$^{112}$Sn at 50 MeV/A energy).  As this involves two
ratios and two different systems, the analysis is considerably more
complicated than what we considered before.  The ratio $R$ we are after
is given by
\begin{eqnarray}
R=\frac{\omega^{(2)}_{l+m,k}}{\omega^{(2)}_{l,k}}
\frac{\omega^{(1)}_{l,k}}{\omega^{(1)}_{l+m,k}}
\frac{Q_{N2-l-m,Z2-k}}{Q_{N2-l,Z2-k}}
\frac{Q_{N1-l,Z1-k}}{Q_{N1-l-m,Z1-k}}
\end{eqnarray}
For central collisions at the same beam energy per particle, the $\omega$
factors will give unity.  Employing the saddle-point approximation
and setting the ratios of the $det$'s as unity as before, we can consider
\begin{eqnarray}
\frac{Q_{N2-l-m,Z2-k}}{Q_{N2-l,Z2-k}}=\frac{e^{-(\lambda_n(N2-l-m)+
\lambda_p(Z2-k))+
logZ_{gr}(\lambda_n,\lambda_p)}}{e^{-(\lambda'_n(N2-l)+\lambda'_p(Z2-k))+
logZ_{gr}(\lambda'_n,\lambda'_p)}}
\end{eqnarray}
Similarly,
\begin{eqnarray}
\frac{Q_{N1-l-m,Z1-k}}{Q_{N1-l,Z1-k}}=\frac{e^{-(\tilde{\lambda}_n(N1-l-m)+
\tilde{\lambda}_p(Z1-k))+
logZ_{gr}(\tilde{\lambda}_n,\tilde{\lambda}_p)}}
{e^{-(\tilde{\lambda}'_n(N1-l)+\tilde{\lambda}'_p(Z1-k))+
logZ_{gr}(\tilde{\lambda}'_n,\tilde{\lambda}'_p)}}
\end{eqnarray}
We can now indicate how the grand canonical results are recovered.  We set
$\lambda_n=\lambda'_n;  \tilde{\lambda}_n=\tilde{\lambda}'_n$ and
$\lambda_n-\tilde{\lambda}_n=\Delta\lambda_n$ then the ratio achieves the
exponential character: $R=\exp(m\Delta\lambda_n)$.  Experimentally
it is found that the relationship $R=\exp(\alpha m)$ where $\alpha$ is
a constant independent of $l$ and $k$ is quite well respected. This is
not so obvious from eqs. (4.1), (4.2) and (4.3).  We are therefore
required to investigate this near independence of the constant $\alpha$.

If we write in Eq.(4.2) $\lambda_n'=\lambda_n+\Delta\lambda_n$ and expand
$logZ_{gr}(\lambda_n',\lambda_p')$ in terms of
$logZ_{gr}(\lambda_n,\lambda_p)$
and keep lowest order corrections, the right hand side of eq.(4.2) is simply
$e^{m\lambda_n'}$.  In a similar fashion, the right hand side of eq.(4.3) is
$e^{m\tilde{\lambda}_n'}$, so that the ratio $R$ of eq.(4.1) is
$\exp(m(\lambda_n'-\tilde{\lambda}_n'))$
where, of course, the values of $\lambda_n',\tilde{\lambda}_n'$ are
chosen to give neutron numbers $N2-l$ and $N1-l$ and proton numbers
$Z2-k$ and $Z1-k$ respectively.  Our
next task is to verify that $\lambda_n'-\tilde{\lambda}_n'$ is approximately
independent of $l$ and $k$.

We have four equations:
\begin{eqnarray}
\sum ie^{i\lambda_n'+j\lambda_p'}\omega_{i,j}=N2-l
\end{eqnarray}
\begin{eqnarray}
\sum je^{i\lambda_n'+j\lambda_p'}\omega_{i,j}=Z2-k
\end{eqnarray}
\begin{eqnarray}
\sum ie^{i\tilde{\lambda}_n'+j\tilde{\lambda}_p'}\omega_{i,j}=N1-l
\end{eqnarray}
\begin{eqnarray}
\sum je^{i\tilde{\lambda}_n'+j\tilde{\lambda}_p'}\omega_{i,j}=Z1-k
\end{eqnarray}
Let $\lambda_n'=\tilde{\lambda}_n'+\delta\lambda_n$ and $\lambda_p'=
\tilde{\lambda}_p'+\delta\lambda_p$.  From Eqs. (4.5) and (4.7), retaining
terms to lowest order in $\delta\lambda_p$ and $\delta\lambda_n$ we obtain
\begin{eqnarray}
\delta\lambda_p\sum
j^2e^{i\tilde{\lambda}_n'+j\tilde{\lambda}_p'}\omega_{i,j}
+\delta\lambda_n\sum
ije^{i\tilde{\lambda}_n'+j\tilde{\lambda}_p'}\omega_{i,j}
=Z2-Z1
\end{eqnarray}
In a similar fashion from Eqs. (4.4) and (4.6) we can obtain
\begin{eqnarray}
\delta\lambda_n\sum
i^2e^{i\tilde{\lambda}_n'+j\tilde{\lambda}_p'}\omega_{i,j}
+\delta\lambda_p\sum
ije^{i\tilde{\lambda}_n'+j\tilde{\lambda}_p'}\omega_{i,j}
=N2-N1
\end{eqnarray}
Eqs (4.8) and (4.9) can be solved for $\delta\lambda_n$ and $\delta\lambda_p$
and in the lowest order these value are independent of $l$ and $k$ but
depend upon $N2, Z2, N1$ and $Z1$.
To this order $R$ of eq.(4.1) is independent of $l$ and $k$ as it
is in the usual grand canonical ensemble.  This is seen to be obeyed in
experiments to a good approximation.

Instead of eqs.(4.4) to (4.9), one may also consider the following
approximation
trying to relate to the grand canonical ensemble.  Recall that $\lambda_n',
\lambda_p'$ are the fugacities of a system which has $N2-l$ neutrons
and $Z2-k$ protons.  If we denote the fugacities of the system which
has $N2$ neutrons and $Z2$ protons by $\Lambda_{N2},\Lambda_{Z2}$ and
employ the same approximate methods used in the discussion leading to
eq.(2.11), we get $\lambda_n'=\Lambda_{N2}+\frac{lC_2-kB_2}{B_2^2-A_2C_2}$
Similarly $\tilde{\lambda}_n', \tilde{\lambda}_p'$
refers to a system which has $N1-l$ neutrons and $Z1-k$ protons.  In an
obvious notation we also get $\tilde{\lambda}_n'=\Lambda_{N1}+\frac
{lC_1-kB_1}{B_1^2-A_1C_1}$.  The quantity of interest is
\begin{eqnarray}
\lambda_n'-\tilde{\lambda}_n'=\Lambda_{N2}-\Lambda_{N1}+\frac{lC_2-kB_2}
{B_2^2-A_2C_2}-\frac{lC_1-kB_1}{B_1^2-A_1C_1}
\end{eqnarray}
Because of cancellations in the above equation, results again approximate
the grand canonical result quite closely.

\section{Summary}
In summary, we have explored several experimental observables which are 
sensitive to the isospin effects in multifragmentation. We find that
the mirror ratios, isoscaling 
and temperatures calculated in canonical model behave similarly as 
those predicted with the grand canonical model with one significant
difference: the temperature deduced from the calculated observables
with the canonical model using the rules based on the grand canonical
model can be significantly different from the true temperatures.

\section{Acknowledgment}
This work is supported in part by the Natural Sciences and Engineering
Research Council of Canada, by {\it le Fonds pour la Formation
de Chercheurs et l'aide \`a la Recherche du Qu\'ebec} and the National
Science Foundation Grant No. PHY-95-28844.

\begin{figure}
\caption{Isobar ratios for three pairs of mirror nuclei obtained from the 
central collisions of $^{124}$Sn+$^{124}$Sn (solid points) and 
$^{112}$Sn+$^{112}$Sn (open points) collisions. The lines are best
fit of Eq. 1.4.}
\end{figure}

\begin{figure}
\caption{Exact Canonical Model calculations for a system of neutron number
N=104 and Z=70 using a freeze-out density of one-quarter of normal density.  
This simulates central $^{124}$Sn+$^{124}$Sn collisions.
Lower values of N and Z used in this calculation reflect effects of 
pre-equilibrium emissions. 
The ratios of yields of mirror nuclei are plotted for $a=1,3,7,9,11$ and 13.
The results for 3,7 and 11 can be compared with the experimental results
(Fig.1). $T_{actual}$=5 MeV is the temperature used in the canonical model 
calculation; $T_{best fit}$ would be the temperature deduced if one 
fit the solid points obtained from the canonical calculations,
using the grand canonical formula, Eq. (1.4).  The
results from a saddle-point Eq. (2.10) approximation are also shown.}
\end{figure}

\begin{figure}
\caption{The inverse Albergo temperature, Eq. (3.2), from the canonical model
is plotted as a function of the binding energy difference, B. 
The inset shows the predicted temperature in an expanded scale. 
The dash line at T=5 MeV is the input temperature to the calculation.}
\end{figure}
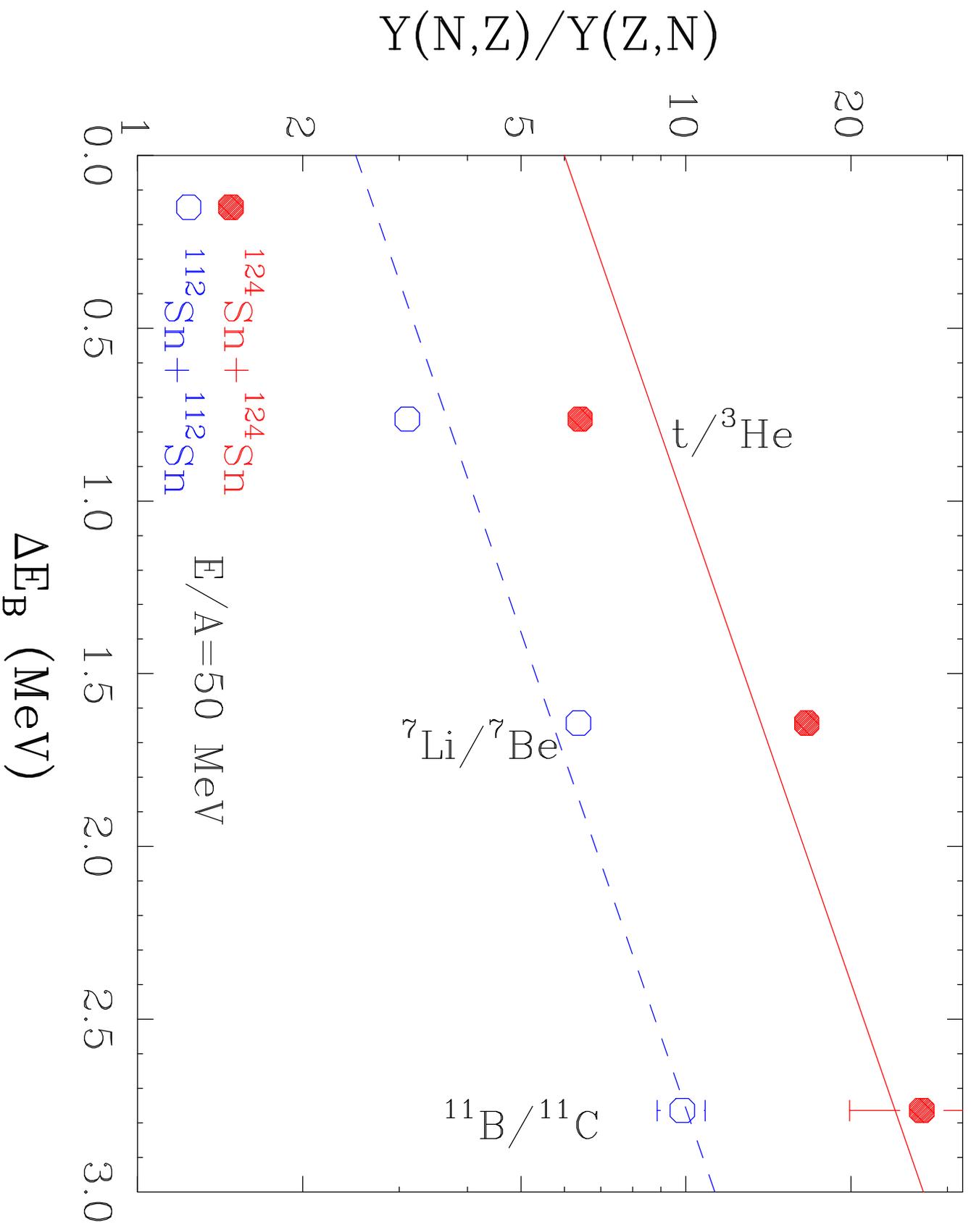
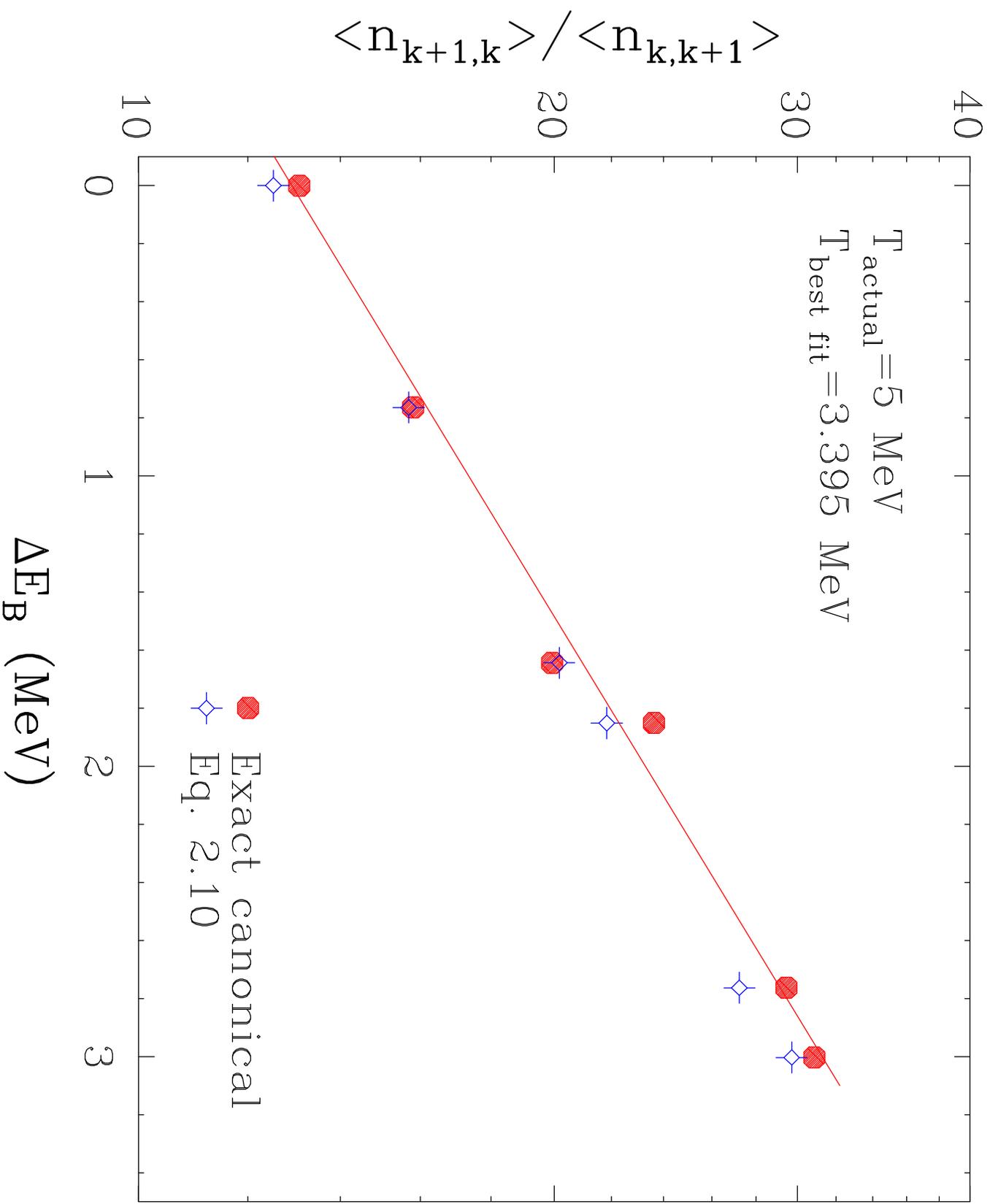
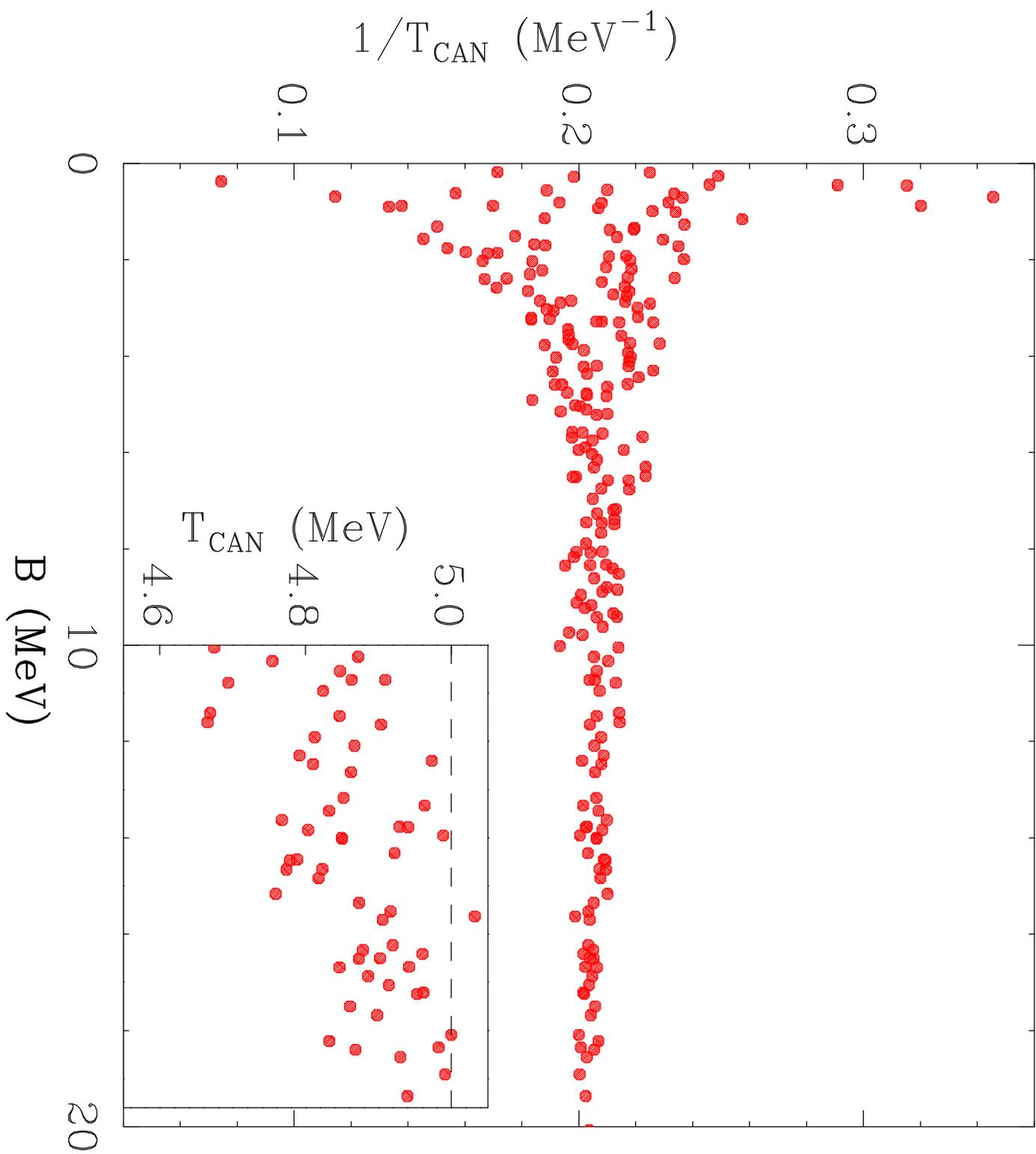
\end{document}